\documentclass[twocolumn,showpacs,aps,floatfix,pra,superscriptaddress,nofootinbib]{revtex4}
\usepackage{graphicx}
\usepackage{bm}
\usepackage{amsmath}
\def\beq{\begin{equation}}
\def\eeq{\end{equation}}
\def\beqa{\begin{eqnarray}}
\def\eeqa{\end{eqnarray}}
\def\la{\langle}
\def\ra{\rangle}
\def\T{\textsf{T}}
\def\sa{\textsf}

\newcommand{\ket}[1]{|#1\rangle}
\begin {document}
\title{Quantum motion effects in an ultracold-atom  Mach-Zehnder interferometer}
\author{I. Lizuain}
\email[Email address: ]{ion.lizuain@ehu.es}
\affiliation{Departamento de Qu\'\i mica-F\'\i sica,
UPV-EHU, Apdo. 644, Bilbao, Spain}
\author{S. V. Mousavi}
\email{s_v_moosavi@mehr.sharif.edu}
\affiliation{Department of Physics, Sharif University of Technology, P. O. Box 11365-9161, Tehran, Iran}
\affiliation{Department of Physics, The University of Qom, P. O. Box 37165, Qom, Iran}
\author{D. Seidel}
\email[Email address: ]{dirk_seidel@ehu.es} 
\affiliation{Departamento de Qu\'\i mica-F\'\i sica,
UPV-EHU, Apdo. 644, Bilbao, Spain}
\author{J. G. Muga}
\email[Email address: ]{jg.muga@ehu.es} 
\affiliation{Departamento de Qu\'\i mica-F\'\i sica,
UPV-EHU, Apdo. 644, Bilbao, Spain} 


\begin{abstract}
We study the effect of quantum motion in a Mach-Zehnder interferometer 
where ultracold, two-level atoms cross a $\pi/2\,$-$\pi\,$-$\pi/2$ configuration  
of separated, laser illuminated regions. 
Explicit and exact expressions are obtained for transmission amplitudes of monochromatic, incident atomic waves using recurrence relations which take into account all possible paths: the direct ones usually considered in the simple semiclassical treatment, but including  
quantum motion corrections, 
and the paths in which the atoms are repeatedly reflected at the fields.       
\end{abstract}
\maketitle
\section{Introduction}
The fringes of an interferometer are sensitive 
to differential phases of the arms caused by   
unequal fields along the interfering paths.  
This makes interferometers useful for metrology and fundamental studies.    
In particular, atom interferometers offer, because of the internal structure of the atom, richer interactions,  
greater and simpler control  
than the ones based on light, electrons or neutrons \cite{ai}. 
They are used for the precise determination of frequencies
and times in atomic clocks, as well as many other applications 
to measure with unprecedented accuracy the gravity field \cite{KaCh-1991} and gravity gradients \cite{ka98}, 
rotations \cite{RiKiWiHeBo-1991,GLK00}, fundamental constants \cite{WeYoCh-1994}, accelerations \cite{chu99}, 
or relativistic effects \cite{rela}. 
Indeed, the accuracy level is currently so high that the theoretical treatments of the 
global performance of the interferometer \cite{AB03,DB06} or the individual constituents
(beam splitters, mirrors) \cite{AB05, Antoine07} need to be refined 
with respect to the simple, original modelings.  
A further reason is the use of ultracold atoms
\cite{WeYoCh-1994,Shi,Nuc}  
to minimize velocity broadening and 
to increase coherence lengths and flight times; they also make possible the spatial separation 
of the arms 
by atomic recoil \cite{Bo-1989, ChDuKaYa-1989}, as in Sagnac interferometry, where    
slower atoms increase arm separation, the area enclosed, and thus the sensitivity achieved.      
In addition, the use of condensates and other ultracold-matter phases (such as the Tonks-Girardeau gas) 
in internal-state interferometry is currently being explored \cite{Band,TG}. 
Since the atomic velocities may be nowadays several orders of magnitude smaller than in 
early beam experiments \cite{Ramsey}, 
these developments raise the following question: 
Is there any fundamental or practical 
lower bound for the velocities in interferometry? \cite{Israel}.  
To answer it we need to go beyond 
the approximation in which the center of mass motion along the interferometer arms is
treated classically.    
\begin{figure}[t]
\begin{center}
\includegraphics[height=3.5cm]{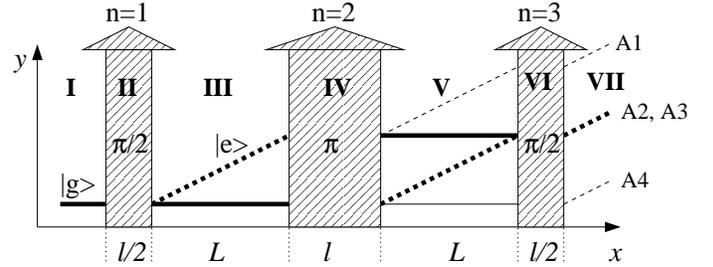}
\caption[]{Atom interferometry with a $\pi/2-\pi-\pi/2$ pulse configuration. The left edges of the three pulses
are located at $x_1=0$, $x_2=L+l/2$ and $x_3=2L+3l/2$.
The beams are not drawn to scale and their width is greatly exaggerated.}
\label{chu_setup_fig}
\end{center}
\end{figure}

Much work on that line by Bord\'e and coworkers has emphasized a wave packet
approach \cite{AB03,AB05}.  
We shall explore here a complementary stationary method, extending some previous results 
on recurrence relations which were applied to Ramsey interferometry in a waveguide \cite{SeiMu-EPJD-2006}. 
The analysis of stationary solutions 
leads to useful insight, and quite frequently provides sufficient information, as the 
history of scattering theory demonstrates (consider, e.g., the cavalier 
but straightforward derivation of cross sections from stationary waves
versus the more rigorous and cumbersome, but finally equivalent, wave packet derivation).
Of course, wave packets can be constructed  
afterwards by linear superposition for examining transients and specific 
space-time  processes.  
Among the interferometers with spatially separated paths we shall 
focus on the simplest configuration, a Mach-Zehnder interferometer,
first implemented in the time domain by Kasevich and Chu \cite{KaCh-1991}. 
We shall deal here with the version in which the laser beams are separated
in space \cite{GLK00}. 
It consists on a first $\pi/2$
laser beam acting as an atom-beam splitter, followed by 
a mirror ($\pi$ beam) and finally a second, recombining $\pi/2$ beam, see Fig. \ref{chu_setup_fig}.  
Our main tool in this investigation is the implementation of exact 
relations for the final transmission amplitude in the excited state. 
They may be cast as ``recurrence relations'' in terms of the 
scattering amplitudes 
for each laser field \cite{Rozman,SeiMu-EPJD-2006}, which allows us to classify      
and calculate all possible paths by the number of reflections \footnote{In this paper the term 
``reflection'' refers to a  
change in the sign of the momentum component in the longitudinal $x$-direction.
Do not confuse this with the recoil
taking place at the second (mirror) laser, in which the excited or ground state are interchanged but the 
momentum component in $x$-direction does not change sign.}: the dominant or ``direct'' ones 
(without reflections),  
associated with the usual semiclassical ordering of events but affected by quantum corrections, 
and also those paths in which the 
particle is reflected in several field regions. 
The extreme low-velocity regime in which these later ``multiple scattering'' 
paths become significant 
distorts severely the interference pattern and thus sets a fundamental lower limit to the atomic velocities for interferometry with fields separated in the space domain \cite{SeiMu-EPJD-2006,Israel}; for intermediate velocities, just above  the multiple scattering regime, 
direct paths dominate, but the semiclassical expressions are not yet
quite accurate and need correction. 
Therefore, an understanding of the various effects and scales involved is useful.   
To simplify the analysis and isolate quantum motion effects from other 
phenomena we shall ignore in this paper 
any external fields different from the laser fields.   
Other simplifying assumptions are the consideration of flat and sharp 
(square) laser sheets, fully coherent processes (i.e., we neglect excited state relaxation), 
and semiclassical atom-laser interaction.   
Some of these approximations are discussed in 
the final section.


\section{Notation and Hamiltonian}
\label{notation_sec}
\subsection{Atom field interaction in 3D}
We consider a setup where a two-level atom,  with an internal (hyperfine) 
transition frequency $\omega_{ge}$ between levels $|g\ra$ and $|e\ra$,
moves with an initial wavenumber $\boldsymbol{k}=(k_x,k_y,k_z)$ ($k_x\gg k_y,k_z$) 
and is illuminated 
in three $x$-localized regions by a classical
electric field  $\boldsymbol{E}(x,t)=\boldsymbol{E_0}(x) \cos {(\omega_L t-k_L y+\phi(x)})$ traveling in $y$-direction, 
see Fig. \ref{chu_setup_fig}. 
The full 3D Hamiltonian describing this system in the Shr\"odinger picture is
\begin{eqnarray}
H&=&\frac{\boldsymbol p^2}{2m}+\hbar\omega_{ge}|e\ra\la e|
\nonumber\\
&+&\hbar\Omega(x)\left(\sigma_++\sigma_-\right)
\cos {\left[\omega_L t-k_L y+\phi(x)\right]},
\end{eqnarray}
where $\sigma_+=|e\ra\la g|$, $\sigma_-=|g\ra\la e|$, 
the $x$-dependent Rabi frequency $\Omega(x)$ is assumed to be 
constant inside the field regions, $\Omega(x) = \Omega$ for $x\in [0,l/2]$, $x\in [L+l/2,L+3l/2]$ and 
$x\in [2L+3l/2,2L+2l]$ and zero otherwise, and  the laser phase 
$\phi(x)$ is constant 
within each of the illuminated regions with values $\phi_{n},\, 
n=1,2,3$. 
In practice the traveling wave is an effective one 
corresponding to two counter propagating lasers which induce a two-photon 
Raman transition and a large (optical) recoil and arm separation, so that   
the parameters are effective ones, after adiabatic elimination of 
a non-resonant upper state, see e.g. \cite{Bordeinai,YouKaChu-in-book}.  
In a field adapted interaction-picture
defined by $H_0=\hbar\omega_L|e\ra\la e|$, and applying the rotating-wave
approximation (RWA),
the time dependence of the Hamiltonian is removed, 
\begin{equation}
H_I^{RWA}=\frac{\boldsymbol p^2}{2m}-\hbar\Delta_0|e\ra\la e|+\frac{\hbar\Omega(x)}{2}
\left[e^{i\left[k_Ly-\phi(x)\right]}\sigma_++H.c\right],
\end{equation}
where $\Delta_0=\omega_L-\omega_{ge}$ is the detuning between the laser frequency and the internal transition.

\subsection{1D effective equation in $x$-direction}
To solve the stationary Shr\"odinger equation 
$H_I^{RWA}|\phi_k(x,y,z)\ra=E_k|\phi_k(x,y,z)\ra$ for an energy $E_k=\hbar^2k^2/(2m)$, 
with $k^2=k_x^2+k_y^2+k_z^2$, we use the ansatz
\begin{equation}
|\phi_k(x,y,z)\ra=g_x(x)e^{ik_yy}e^{ik_zz}|g\ra+e_x(x)e^{iq_yy}e^{ik_zz}|e\ra,
\end{equation}
which describes the momentum transfer in $y$-direction when the atom 
is excited,  $q_y=k_y+k_L$, 
and conservation of momentum (free evolution) in $z$-direction. 
Inserting this 
ansatz into the Shr\"odinger equation gives an effective equation in $x$ direction,
\beq
H_x\binom{g_x}{e_x}=E_x\binom{g_x}{e_x},
\label{sef}
\eeq
where $E_x=\frac{k_x^2\hbar^2}{2m}$, 
\begin{equation}
\label{effective_x_ham}
H_x=\frac{p_x^2}{2m}-\hbar\Delta|e\ra\la e|+\frac{\hbar\Omega(x)}{2}
\left[e^{-i\phi(x)}\sigma_++H.c\right],
\end{equation}
and $\Delta$ is the effective detuning, 
\begin{eqnarray}
\Delta&=&\Delta_0-\Delta_{kin},
\label{det1}\\
\Delta_{kin}&=&\frac{\hbar k_L^2}{2m}+\frac{\hbar k_yk_L}{m},
\label{det2}
\end{eqnarray}
which includes the ordinary detuning $\Delta_0$ 
and the ``kinetic detuning'' $\Delta_{kin}$ 
with a photon recoil term 
and a Doppler term.
From now on we shall deal with the 1D effective equation (\ref{sef}) only.    

\section{Semiclassical regime}
\label{semiclassical_sec}
In the simplest treatment, valid for fast enough particles, $E_x\gg\hbar\Omega$,
$k_x\approx q_x\equiv\sqrt{k_x^2+2m\Delta/\hbar}$, 
the internal and longitudinal degrees of freedom are decoupled,
and the  
$x$-component of the center of mass is assumed to follow the classical trajectory 
$x(t)=v_x t$, where $v_x=\hbar k_x/m$. We shall, 
in other words, treat the interferometer with fields
separated in space as an interferometer for fields separated in time. 
Then, the internal states will evolve with $H_n$,
\begin{equation}
H_n=-\hbar\Delta|e\ra\la e|+\frac{\hbar\Omega}{2}
\left(e^{-i\phi_{n}}\sigma_++H.c\right),
\label{Hn}
\end{equation}
in the $n$th laser field, and with the bare Hamiltonian 
\begin{equation}
H_B=-\hbar\Delta|e\ra\la e|
\end{equation}
in the non-interacting regions.
The corresponding time evolution operators are $e^{-iH_n t/\hbar}$ in the $n$th 
laser and $e^{-iH_Bt/\hbar}$ in the free evolution regions.  
From Fig. \ref{chu_setup_fig} it can be seen that there are four possible 
semiclassical paths which lead to an excited atom from an atom which is initially in the ground 
state $|g\ra$. If we set $\tau=l/v_x$ and $T=L/v_x$,
the amplitudes of these four paths are given by
\begin{widetext}
\begin{eqnarray}
A_1^{scl}&=&\la e|e^{-iH_3\tau/(2\hbar)}e^{-iH_BT/\hbar}|e\ra   
\la e|e^{-iH_2\tau/\hbar}e^{-iH_BT/\hbar}|e\ra
\la e|e^{-iH_1\tau/(2\hbar)}|g\ra,
\nonumber\\
A_2^{scl}&=&\la e|e^{-iH_3\tau/(2\hbar)}e^{-iH_BT/\hbar}|g\ra
\la g|e^{-iH_2\tau/\hbar}e^{-iH_BT/\hbar}|e\ra
\la e|e^{-iH_1\tau/(2\hbar)}|g\ra,
\nonumber\\
A_3^{scl}&=&\la e|e^{-iH_3\tau/(2\hbar)}e^{-iH_BT/\hbar}|e\ra   
\la e|e^{-iH_2\tau/\hbar}e^{-iH_BT/\hbar}|g\ra   
\la g|e^{-iH_1\tau/(2\hbar)}|g\ra,
\nonumber\\
A_4^{scl}&=&\la e|e^{-iH_3\tau/(2\hbar)}e^{-iH_BT/\hbar}|g\ra   
\la g|e^{-iH_2\tau/\hbar}e^{-iH_BT/\hbar}|g\ra
\la g|e^{-iH_1\tau/(2\hbar)}|g\ra,
\label{as}
\end{eqnarray}
\end{widetext}
For a wave-packet,
the momentum recoil will lead to a separation of these paths in the $y$-direction. If this 
separation is larger than the transversal position spread of the packet $\Delta y$, $\hbar k_LT/(2m)\gg\Delta y$, 
and the detector's resolution is better than $\hbar k_LT/(2m)$,
the interference between the outer paths will be suppressed  and only the interference between $A_2^{scl}$ and $A_3^{scl}$ will be observed, see Fig. \ref{chu_setup_fig} (thick lines). 
The corresponding excitation probability is
(see the Appendix A for explicit expressions of the matrix elements in  
Eq. (\ref{as}))
\begin{eqnarray}
\label{pge_scl_eq}
P_{ge}^{scl}&=&\left|A_2^{scl}+A_3^{scl}\right|^2
=\frac{\Omega^2}{4\Omega'^6} \sin^2 \frac{\Omega'\tau}{2}  \nonumber\\
&\times&
\Bigl \{ 4\Delta^2 \Omega'^2 + 3\Omega^4 + \Omega^2 \Bigl[ 4\Delta^2 
\cos \frac{\Omega'\tau}{2} + \Omega^2 \cos (\Omega'\tau)  \nonumber\\ 
&-& 4 \Bigl ( \Delta^2 + \Omega'^2 + \Omega^2 \cos \frac{\Omega'\tau}{2} \Bigr ) 
\sin^2 \frac{\Omega'\tau}{4}\cos \Phi  \Bigr] \Bigr \},
\end{eqnarray}
where $\Phi$ is a combination of the three individual laser phases, $\Phi=\phi_1-2\phi_2+\phi_3$,
and $\Omega'=\sqrt{\Omega^2+\Delta^2}$, see Fig. \ref{shift_fig} (solid line).

Remark 1: Unlike the Ramsey configuration, this pattern is independent of $T$ and thus of the intermediate distance $L$ between the pulses since  
both interfering paths
spend the same amount of time in the upper level and there is no accumulation of a phase difference.    

Remark 2: $dP_{ge}^{scl}/d\Phi\propto\sin\Phi$ so that there is a minimum (zero) at $\Phi=0$ 
independently of all other parameters. In particular, this is true 
(within this approximation) regardless of the 
velocity and 
the precision with which the $\pi/2$ and $\pi$ pulses are implemented.    

Remark 3: The detuning
may affect the visibility of the fringes but does not
shift the fringe pattern.
Near resonance, $\Delta\ll\Omega$, and considering perfect pulse areas, $(\Omega\tau=\pi)$, 
Eq. (\ref{pge_scl_eq}) (which does not depend on these conditions) 
may be expanded to leading order in $\Delta$ as
\begin{eqnarray}
\label{smcl_ampl_det}
P_{ge}^{scl}&\approx&
\sin^2\frac{\Phi}{2}\left(1-\frac{\Delta^2}{\Omega^2}\right).
\end{eqnarray}
The semiclassical central zero is in summary robust versus velocity or detuning variations, 
and does not require strict ($\pi/2$ and $\pi$)
conditions on the pulse areas.      
However, if the kinetic energy of the atom is comparable with the interaction energy, this 
semiclassical approach breaks down and a full quantum mechanical solution becomes necessary,
yielding a phase shift of the interference pattern as we shall see.
%
\section{Quantum treatment}
\label{quantum_sec}
The general solution to the stationary Schr\"odinger 
equation (\ref{sef}) away from the laser fields takes the form 
\begin{eqnarray}
\label{asymptotic_solution}
|\psi_x\ra&=&\left(G_+ e^{ik_xx}+G_- e^{-ik_xx}\right)|g\ra
\nonumber\\
&+&\left(E_+ e^{iq_xx}+E_- e^{-iq_xx}\right)|e\ra,
\end{eqnarray}
where
the amplitudes $G_\pm$ and $E_\pm$ have to be determined from  the boundary and matching conditions. 
We will follow Ref. \cite{SeiMu-EPJD-2006} to derive the exact quantum result of the interference pattern. 
Let us denote by $R_{ij}^{l}$ ($T_{ij}^{l}$) the total reflection (transmission) 
amplitudes of an atom entering the interferometer from the left in the $i$ channel and an outgoing plane wave in the 
$j$ channel, $i,j=g,e$. In the case of a plane wave incident from the left in the ground state, we have, for the leftmost laser-free region I, see Fig. \ref{chu_setup_fig},     
\begin{equation}
|\psi\ra_{_I}=\left(e^{ik_xx}+R_{gg}^l e^{-ik_xx}\right)|g\ra+R_{ge}^le^{-iq_xx}|e\ra,
\end{equation}
whereas the outgoing wave to the right of field 3, region $VII$, has the form
\begin{equation}
|\psi\ra_{_{VII}}=T_{gg}^l e^{ik_xx}|g\ra+T_{ge}^l e^{iq_xx}|e\ra.
\end{equation}
That is, after passing the three laser pulses the atom may still be in the ground state, propagating with a wavevenumber 
$k_x$, or in the excited state, propagating with a wavenumber $q_x$.
In the latter case, the atomic transition $\ket{g} \to \ket{e}$ induced by the laser field changes the kinetic energy in the effective equation for $x$-direction. 
For $\Delta>0$ the kinetic energy of the excited state component is enhanced by $\hbar\Delta$ whereas for $\Delta <0$ it is reduced by $\hbar\Delta$. For $\Delta$ smaller then the critical value $\Delta_{cr}=-\hbar k_x^2/2m$, the excited state component becomes evanescent and its transmission probability vanishes (the channel becomes closed). Thus, the quantum mechanical probability to observe the transmitted atom in the excited state is zero for $\Delta \leq \Delta_{cr}$; otherwise
\begin{eqnarray}
\label{quantum_pge}
P_{ge}^q=\frac{q_x}{k_x}\left|T_{ge}^l\right|^2 &\textrm{for}& \Delta>\Delta_{cr}, 
\end{eqnarray}
and we shall limit the analysis to this later case. (For a study of the evanescent regime see 
\cite{ROS}). 
The exact form of $T_{ge}^l$ follows from the matching conditions between the free-space solutions and the dressed state solutions inside the fields
using the transfer matrix formalism \cite{fl,SeiMu-EPJD-2006,DaEgHeMu}.
%
%
\subsection{Excited state probability amplitude}
The solutions (\ref{asymptotic_solution}) in the laser free regions may be given in a compact form by 
a constant $4$-dimensional vector $\boldsymbol{v}=(G_+,G_-,E_+,E_-)'$ (the prime means ``transpose'') with the complex amplitudes. 
In particular, the scattering boundary conditions are imposed on the external regions 
$I$ and $VII$,  
\begin{eqnarray}
\boldsymbol{v}_{_{I}}&=&(1,R_{gg}^l,0,R_{ge}^l)',\\
\boldsymbol{v}_{_{VII}}&=&(T_{gg}^l,0,T_{ge}^l,0)'.
\end{eqnarray}
These solutions at the external regions are related by a combination of transfer matrices \cite{SeiMu-EPJD-2006,DaEgHeMu}, 
see Appendix \ref{transfer_matrix},  
as
\begin{equation}
\label{vin_T_vout}
\boldsymbol{v}_{_{I}}=\T^{(1)}\T^{(2)}\T^{(3)}\boldsymbol{v}_{_{VII}}=\T^{tot}\boldsymbol{v}_{_{VII}},
\end{equation}
where $\T^{(n)}$ is the transfer matrix for the $n$th laser, i. e., 
\begin{eqnarray}
\T^{(1)}&=&\T(0,l/2,\phi_1),
\nonumber\\
\T^{(2)}&=&\T(L+l/2,L+3l/2,\phi_2),
\nonumber\\
\T^{(3)}&=&\T(2L+3l/2,2L+2l,\phi_3), 
\end{eqnarray}
and $\T^{tot}$ is the ``total'' transfer matrix for the whole 
interferometer. 
Solving 
Eq. (\ref{vin_T_vout}) for $T_{ge}^l$, we find that the excited state transmission probability amplitude is given by
\begin{equation}
\label{Tge_eq}
T_{ge}^l=\frac{\T_{31}^{tot}}{\T_{13}^{tot}\T_{31}^{tot}-\T_{11}^{tot}\T_{33}^{tot}}.
\end{equation}
%
%
\subsection{Two-channel recurrence relations}
Since the transfer matrices for the laser interactions are known
in terms of the laser parameters, see Appendix \ref{transfer_matrix},  
Eq. (\ref{Tge_eq}) provides an explicit, and easy to calculate expression. However, 
this numerical calculation alone does not necessarily provide much physical insight.  
It is useful to relate the transfer matrices to scattering
transmission and reflection amplitudes for the individual laser regions. 
We denote by $(r_n)_{ij}^l$ and $(t_n)_{ij}^l$ the single-laser reflection and transmission amplitude for incidence on the $n$th laser ``barrier'' from the left in the $i$th 
channel and an outgoing plane wave in the $j$th channel 
(as before $i,j=g,e$), and by $(r_n)_{ij}^r$ and $(t_n)_{ij}^r$ the
corresponding amplitudes for right incidence.
These scattering  amplitudes for the laser units are also easy to calculate 
(exactly, using Eqs. (\ref{tr1},\ref{tr2},\ref{tram}), or 
with approximations, e.g. semiclassically \cite{fl}  
or otherwise), and their moduli are typically close to one or zero,
so that we may introduce an expansion parameter, see below, to discern
the dominant contributions corresponding to ``direct scattering'',  and classify 
the order of the corrections in terms of the number of reflections.      
One further advantage is that we may also classify 
and distinguish the paths according to the number of inter-laser 
free-motion regions in which the atom flies in the excited state
(for direct, reflectionless paths, this number may be $0$, as in A4 of
Fig. \ref{chu_setup_fig}, $1$, as in A2 and A3, or $2$, as in A1).  
We may thus distinguish those paths that will finally interfere (e.g., A2 and A3 in Fig. \ref{chu_setup_fig})
from those that will not (A1 and A4 in Fig. \ref{chu_setup_fig}) in an atomic wave-packet  
because of the arm separation due to recoil. 
 
The relation between the matrix elements $\T^{(n)}_{ij}$ and
the individual scattering amplitudes $(r_n)_{i,j}^{l,r},(t_n)_{i,j}^{l,r}$ is invertible, i. e., 
$\T^{(n)}_{ij}=f_n\left[(r_n)_{i,j}^{l,r},(t_n)_{i,j}^{l,r}\right]$
with invertible known functions $f_n$ (see Appendix \ref{recurrence}).
%
\subsection{Mach-Zehnder terms}
If the kinetic energy $(\hbar k_x)^2/(2m)$ is larger than the Rabi
energy $\hbar\Omega$ the scattering process will be dominated by transmission through all fields and all reflection amplitudes will be small quantities compared to the transmission ones, i. e., 
$|(t_n)_{ij}^{l,r}| \gg |(r_n)_{ij}^{l,r}|$ for all lasers ($n=1,2,3$). 
Moreover, the second laser
is assumed to apply very nearly a $\pi$-pulse which flips the internal atomic state, so that 
$|(t_2)_{ii}^{l,r}|\ll1$. Multiplying all small 
amplitudes by a small expansion parameter $\eta$, 
the series expansion in $\eta$ of $T_{ge}^l$ has the form
\begin{equation}
\label{tge_26paths}
T_{ge}^l=\left(A_2^q+A_3^q\right)+
\eta \left(A_1^q+A_4^q\right)+
\eta^2 \sum_{i=1}^{22} B_i^q, 
\end{equation}
where the individual quantum amplitudes $A_ i^q$ and $B_i^q$ are given in 
Appendix \ref{explicit_ampl_appendix} in terms
of the single laser scattering amplitudes $(r_n)_{ij}^{r,l}$ and $(t_n)_{ij}^{r,l}$.

The dominant zeroth order terms are in correspondence 
with the $A_2^{scl}$ and $A_3^{scl}$
direct paths in the semiclassical picture, whereas 
first order corrections correspond to the semiclassical $A_1^{scl}$ and $A_4^{scl}$ 
paths and are small since they
contain diagonal transmission amplitudes in the second laser (Appendix \ref{4_paths_app}). 
The quantum reflection effects are included in the second order terms, 
which contain the quantum amplitudes $B_i^q$ 
for all the $22$ possible paths leading to a transmitted
excited atom including two reflections, 
see Appendix \ref{22_paths_app}. 

\begin{figure}[t]
\begin{center}
\includegraphics[height=6cm]{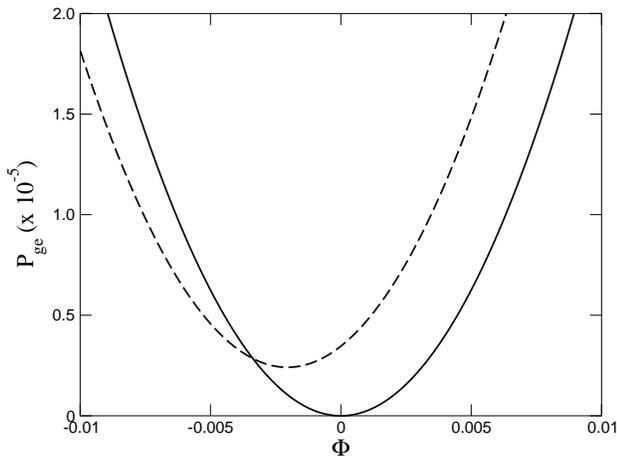}
\caption[]{$P_{ge}^{scl}$ (Eq. (\ref{pge_scl_eq}), solid line) and $P_{ge}^{q}$ 
(Eq. (\ref{pge_quantum_eq}), dashed line) as a function of the phase difference $\Phi$ for a resonant
condition $\Delta=0$. Data: $v_x=1$ cm/s, $l=10\mu$m, $m_{Na}=3.82\times10^{-26}$ Kg. 
The Rabi frequency is fixed to satisfy the $\pi$-pulse condition, $\Omega=\pi v_x/l=2\pi\times 500$ Hz.}
\label{shift_fig}
\end{center}
\end{figure}

In a wave packet, recoil effects will separate in space all these paths 
leading to transmitted excited atoms. 
In order to compare the quantum interference pattern with the one obtained semiclassically in Eq. (\ref{pge_scl_eq}), 
we choose those paths interfering with $A_2^q$ and $A_3^q$, i.e., the ones 
in the Mach-Zehnder geometry (along thick lines in Fig. \ref{chu_setup_fig}): 
they are $B_i^q$ with $i=1,2,3,4$.  
This gives the following transmission amplitude,
\begin{equation}
T_{ge}^l\approx A_2^q+A_3^q +\sum_{i=1}^4B_i^q,
\end{equation}
and the excited state probability
\begin{equation}
P_{ge}^q\approx\frac{q_x}{k_x}\left|A_2^q+A_3^q +\sum_{i=1}^4B_i^q\right|^2,
\label{pge}
\end{equation}
see Eq. (\ref{quantum_pge}). 
\subsection{Direct scattering and quantum shifts}
\begin{figure}[t]
\begin{center}
\includegraphics[height=6cm]{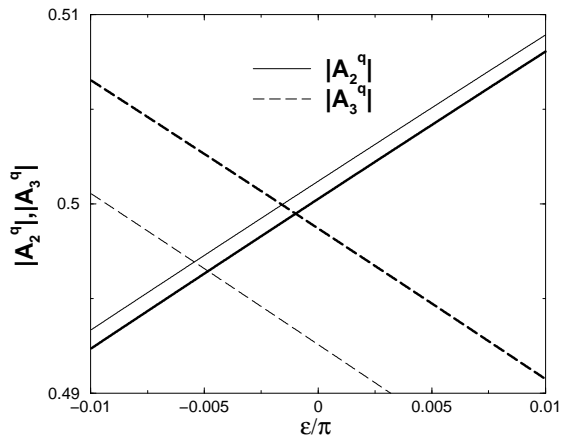}
\caption[]{$|A_2^q|$ and $|A_3^q|$ as a function of $\epsilon$ (the deviation from perfect pulse areas, i. e., 
$\Omega l/v_x=\pi+\epsilon$) for different velocities: The thin lines correspond to $v_x=0.5$ cm/s and the thick ones to $v_x=1.0$ cm/s ($l=10\, \mu$m in both cases). Note that the two moduli cross each other at some value of $\epsilon$. At these values,
$|A_2^q|=|A_3^q|$ and maximum visibility will be obtained, see Eq. \ref{a2a3}.}
\label{moduli}
\end{center}
\end{figure}

\begin{figure}[t]
\begin{center}
\includegraphics[height=6cm]{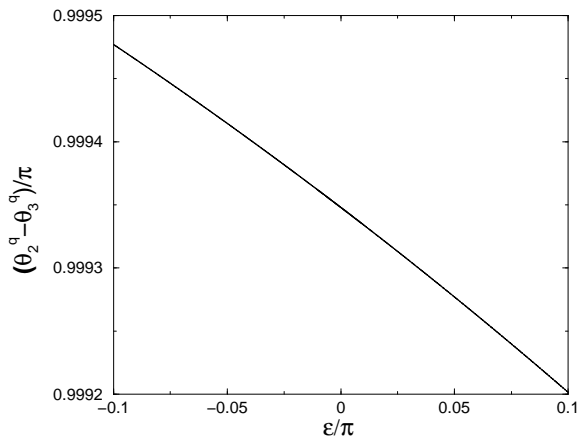}
\caption[]{Quantum phase difference for non-perfect pulse areas for fixed values of  $v=1$ cm/s and $l=10\, \mu$m. Note that
in the semiclassical case this difference is $\theta_2^{scl}-\theta_3^{scl}=\pi$ for every value of $\epsilon$, i.e, for 
every pulse length. }
\label{phase}
\end{center}
\end{figure}

We shall first work out the direct scattering case in which the reflection terms can be neglected,  
\begin{equation}
P_{ge}^q\approx\frac{q_x}{k_x}\left|A_2^q+A_3^q \right|^2. 
\end{equation}
Let us first write the amplitudes in terms of their moduli and phases,   
\beqa
A_2^q&=&(t_1)_{ge}^l (t_2)_{eg}^l (t_3)_{ge}^l
\\
&=&|(t_1)_{ge}^l(t_2)_{eg}^l (t_3)_{ge}^l|
e^{-i(\phi_1-\phi_2+\phi_3)}e^{i\theta_2}
\nonumber\\
A_3^q&=&(t_1)_{gg}^l (t_2)_{ge}^l (t_3)_{ee}^l
=|(t_1)_{gg}^l (t_2)_{ge}^l (t_3)_{ee}^l|
e^{-i\phi_2}e^{i\theta_3}
\nonumber
\eeqa
where the $\theta_{2,3}$ transmission phases 
result from the addition of the phases of the 
individual transmission 
amplitudes along the path when all $\phi_n=0$.  
From the semiclassical expressions, 
\beqa
A_2^{scl}&=&|A_2^{scl}|
e^{-i(\phi_1-\phi_2+\phi_3)}e^{i\theta_2^{scl}},
\\
A_3^{scl}&=&=|A_3^{scl}|
e^{-i\phi_2}e^{i\theta_3^{scl}},
\eeqa
we have that, 
in a $\pi/2-\pi-\pi/2$ configuration with $\Delta=0$, $\theta^{scl}_2-\theta^{scl}_3=\pi$, 
see Appendix A.  For the quantum case 
we may also expect $\delta\theta\equiv\theta^q_2-\theta^q_3\approx\pi$.  
If we write the actual phase as $\delta\theta=\pi+\delta \Phi$,    
there will be a minimum of 
\beq
|A^q_2+A^q_3|^2=|A^q_2|^2+|A^q_3|^2+2\cos(-\Phi+\delta\theta)|A_2^q||A_3^q|,
\label{a2a3}
\eeq
at $\Phi=\delta \Phi$, 
\begin{equation}
\frac{dP_{ge}^{q}}{d\Phi}\propto\sin{\left(-\Phi+\pi+\delta \Phi\right)}=0.
\end{equation}
This is a quantum phase shift which vanishes in    
the semiclassical limit. 
Imposing the condition $\Omega l/v_x=\pi$ at each velocity 
(i.e., exact {\it semiclassical} $\pi/2$ and $\pi$ conditions), 
the quantum motion shift is shown in Fig. \ref{shift_fig} (dashed line).
Take note that the minimum of $P^q_{ge}$ is not a zero  
since the quantum moduli $|A^q_2|$
and $|A^q_3|$ do not exactly
coincide: these conditions do not really split 
the beam in two equal halves at the external lasers, so that  
$|(t_1)^l_{gg}|\ne|(t_1)^l_{ge}|$,  and $|(t_3)^l_{ge}|\ne|(t_3)^l_{ee}|$.
The consequence is a quantum reduction of visibility. 
We may look for enhanced visibility modifying the laser intensity 
and thus the pulse area away from the former condition, i.e.,    
$\Omega l/v_x=\pi+\epsilon$. 
Figs. \ref{moduli} and \ref{phase} show the moduli $|A^q_{2}|$ and $|A^q_3|$ and the
phase shift $\delta \Phi$ as a
function of the extra phase $\epsilon$ for fixed laser width and velocity.
Note that the moduli of $|A^q_2|$ and $|A^q_3|$ cross each other at a value $\epsilon_o$,
%
so that a zero of $P^q_{g,e}$ can indeed be achieved by adjusting the Rabi 
frequency at $(\pi+\epsilon_o)v_x/l$. There are 
however two important differences with respect to the semiclassical 
exact $\pi/2-\pi-\pi/2$ case: (a) the ``optimal'' value $\epsilon_o$ depends on the velocity
(in the semiclassical case $\epsilon_o=0$ for all $v_x$); (b) even for the optimal 
$\epsilon_o$, there is a quantum phase shift, $\delta \Phi\ne 0$.      
\subsection{Reflections} 
\begin{figure}[t]
\begin{center}
\includegraphics[height=6cm]{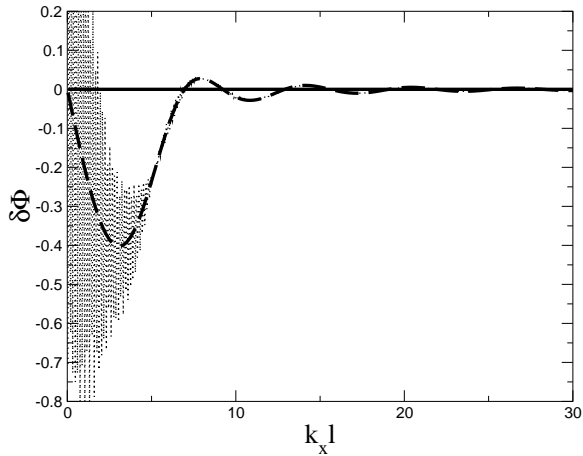}
\caption[]{Phase shift of the interference pattern as a function of $k_x l$. The lenght $l$ is 
kept constant, while the velocity (and thus the Rabi frequency) are changed in order to maintain the $\pi/2$-pulse
condition. It should be clear from Eq. (\ref{pge_scl_eq}) that in the semiclassical regime, there is no phase shift 
(solid line). The dashed line corresponds to the direct scattering approximation, where quantum reflections are 
neglected. 
At low velocities, quantum reflections become relevant and direct approximation breaks down (dotted line). 
Resonant $\Delta=0$ pulses have been considered for the calculations, but numerical simulations show the robustness
of the phase shift against detuning fluctuations. Changes in the detuning of the order $\pm 2v_x/L$ are indistinguishable 
in the scale of the figure, see Sec. \ref{detuning_effect}.}
\label{shift_kx_fig}
\end{center}
\end{figure}

Adding the next order in Eq. (\ref{pge}),      
and taking into account the phase dependence of each of the quantum scattering amplitudes
(Appendix \ref{phase_dep_app}), one may write
\begin{equation}
\label{pge_quantum_eq}
P_{ge}^q\approx\frac{q_x}{k_x}\left|e^{-i\Phi}\left(\tilde A_2^q+\tilde B_2^q+\tilde B_4^q\right)+
\left(\tilde A_3^q+\tilde B_1^q+\tilde B_3^q\right)  \right|^2,
\end{equation}
where the tildes are the amplitudes for $\phi_n=0,\, n=1,2,3$. 
$P_{ge}^q$ is plotted in Fig. \ref{shift_fig} for 
$\Omega l/v_x=\pi$.  
For a velocity region the result cannot be distinguished from 
the calculation with direct paths only, which should be dominant for $\hbar^2 k_x^2/2m\gg \hbar \Omega$. 
Combining this with the $\pi$-pulse condition, the direct scattering
approximation is valid when $k_x l\gg 2\pi$, as it is observed in Fig. \ref{shift_kx_fig},
where, for lower values of
$k_x l$ the direct approximation breaks down and quantum reflection effects become
relevant. The effect is a rather chaotic oscillation of the shift (the actual structure
is even more complex than the one shown in the scale of the figure).  
There are however several reasons why this regime will be difficult to see 
in practice as commented in the final discussion. 
%
%
\subsection{Effect of the detuning}
\label{detuning_effect}
The calculations so far have been made for perfectly resonant interactions, i. e.,
for $\Delta=0$, where the detuning $\Delta$
contains both the natural detuning $\Delta_0$ and the kinetic detuning,
see Eqs. (\ref{det1},\ref{det2}).  
%
%
In a wave packet it is not possible to fulfill the perfect
resonant $\Delta=0$ condition exactly for all components: Even though one may adjust the laser 
frequency to compensate for the recoil term in Eq. (\ref{det2}), 
the momentum spread of the wave-packet in the $y$ direction,  $\Delta k_y$, 
will lead to a  
detuning spread from the Doppler term, $\Delta_D=\hbar k_y k_L/m$. 
In the semiclassical case, we have already shown that the detuning can only affect the visibility of the interference
fringes but will not affect their position, see Eq. (\ref{smcl_ampl_det}).
This is no longer true in 
a fully quantum calculation. 
Since the position spread in the $y$ direction, $\Delta y$, cannot be larger than $\hbar k_L T/2m$  in order to suppress the interference with outer paths, $\Delta k_y$ will also be limited. 
We may thus estimate the Doppler-detuning spread as
\begin{equation}
\Delta_D^{\pm}\approx\frac{\hbar k_L(\Delta k_y)}{m}\approx \pm\frac{2}{T}\approx \pm\frac{2v_x}{L}.
\end{equation}
Numerical simulations with 
$v_x\approx 1$ cm/s and $l=0.1-1.0$ m  show that
a kinetic detuning like this has negligible effect in the calculated phase shift,
which is quite robust against detuning fluctuations.

\section{Discussion}
In this paper we have explored the low velocity limit of atomic interferometry in 
a simple Mach-Zehnder $\pi/2-\pi-\pi/2$ configuration of spatially 
separated laser fields ignoring further external fields. 
In particular, we have performed a fully quantum analysis
of incident monochromatic stationary atomic waves by providing explicit expressions for
transmission probabilities 
from which the physically relevant paths and contributions in terms
of transmission and reflection 
amplitudes for the individual laser fields may be extracted.    

For laser fields separated in space, 
the ideal $\pi/2-\pi-\pi/2$ conditions
leading to perfect splitting, perfect reflection, and 
interferometer phase given exclusively by the laser field phases
cannot be reached in a fully quantum scenario,
even for a fixed incoming velocity. 
The consequence is a quantum-motion phase shift at low atomic velocities 
related to the phases of
the transmission amplitudes. One may optimize the fringe visibility 
by deviating the Rabi frequency from the semiclassical value, 
but a phase shift remains which, in addition, depends on the incident
velocity. This quantum-motion shift is quite insensitive to the 
detuning to be found in wave packet components but shows wild oscillations when 
the velocities are so low that paths with reflections at the fields become
significant. 
  
All the above has been done for square laser profiles in the longitudinal direction, 
with two-channel recurrence relations which are by construction well adapted
to generalizations for more realistic laser intensity profiles.   
We have also considered flat laser sheets ignoring the curvature of the field.  
For direct paths (transmitted in all lasers) this is a good approximation,
whereas paths with reflection, having longer flights and more collisions
with the laser fields, will be more affected by curvature effects, 
which, together with other averaging effects (because of their 
extreme sensitivity to tiny velocity variations)
will surely cancel their contribution to the shift.         
        
\begin{acknowledgments}
We are grateful to M. Kasevich for providing useful information.  
This work has been supported by MEC
(FIS2006-10268-C03-01) and UPV-EHU (00039.310-15968/2004).
D. S. acknowledges a fellowship within the Postdoc-Programme of 
the German Academic Exchange Service (DAAD). 
S. V. M. acknowledges a research visitor Ph. D. student fellowship  
by the Ministry of Science, Research and Technology of Iran.
\end{acknowledgments}
%
%
%
%
%
\appendix
\section{Amplitudes for atom at rest}
These are the amplitudes needed for calculating matrix elements in 
Eq. (\ref{as}) with the Hamiltonian (\ref{Hn}).
Here the atom is illuminated during a time $t$.  
\begin{eqnarray}
\la g|e^{-iH_nt/\hbar}|g\ra&=&e^{i\Delta t/2}[\cos(\Omega't/2)
-i\frac{\Delta}{\Omega'}\sin(\Omega't/2)],
\nonumber\\
\la e|e^{-iH_nt/\hbar}|e\ra&=&e^{-i\Delta t/2}[\cos(\Omega't/2)
+i\frac{\Delta}{\Omega'}\sin(\Omega't/2)],
\nonumber\\
\la e|e^{-iH_nt/\hbar}|g\ra&=&
-ie^{i\Delta t/2}e^{-i\phi_n}\frac{\Omega}{\Omega'}\sin(\Omega't/2),
\nonumber\\
\la g|e^{-iH_nt/\hbar}|e\ra&=&
-ie^{i\Delta t/2}e^{i\phi_n}\frac{\Omega}{\Omega'}\sin(\Omega't/2),
\label{mael}
\end{eqnarray}
where
$\Omega'=\sqrt{\Omega^2+\Delta^2}$. 
They are also useful to obtain semiclassical approximations of matrix elements of 
transmission amplitudes $t_n$. 
\section{Transfer Matrices}
\label{transfer_matrix}
Consider the regions $\alpha=I, II, III$ in Fig. \ref{chu_setup_fig},  
separated by $x_1$ and $x_2$. 
The general solution to the stationary Schr\"odinger equation 
(\ref{sef}) of the effective Hamiltonian (\ref{effective_x_ham}) 
%
%
reads
\begin{equation}
|\psi(x)\ra_\alpha=g_\alpha(x)|g\ra+e_\alpha(x)|e\ra.
\end{equation}
We want to find these solutions for $\alpha=I,II,III$ 
and match them at the boundaries.
%
\subsection{Solution outside and inside the fields}
The solutions at the laser-free regions ($\alpha=I,III$) are given by
\begin{eqnarray}
|\psi(x)\ra_\alpha&=&\left(a_\alpha e^{ik_xx}+b_\alpha e^{-ik_xx}\right)|g\ra 
\nonumber\\
&+&\left(c_\alpha e^{iq_xx}+d_\alpha e^{-iq_xx}\right)|e\ra,
\end{eqnarray}
where $\hbar k_x$ is the initial momentum of the atom in the longitudinal $x$ direction and 
$q_x^2=k_x^2+2m\Delta/\hbar$. 
Inside the laser fields the (unnormalized) dressed state basis which
diagonalizes the interaction part of the Hamiltonian is given by $|\lambda_\pm\ra=|g\ra+2\lambda_\pm e^{-i\phi}\Omega^{-1}|e\ra$, where $\lambda_\pm=(-\Delta\pm\Omega')/2$ are the dressed energies. 
The solution inside the interaction region ($\alpha=II$) will be given in terms of these dressed states and dressed energies,
\begin{eqnarray}
|\psi(x)\ra_{_{II}}&=&\left(a_{II} e^{ik_+x}+b_{II} e^{-ik_+x}\right)|\lambda_+\ra,
\nonumber\\
&+&\left(c_{II} e^{ik_-x}+d_{II} e^{-ik_-x}\right)|\lambda_-\ra,
\end{eqnarray}
with wavenumbers $k_\pm^2=k_x^2-2m\lambda_\pm/\hbar$. The solution in each zone can be then given by a set of
$4$ unknown complex amplitudes, collected in a constant complex vector 
$\boldsymbol{v}_\alpha=(a_\alpha,b_\alpha,c_\alpha,d_\alpha)'$, where the prime means ``transpose''.
\subsection{Matching Conditions: one laser}
The wave functions and their derivatives with respect to $x$ may be written in the following way in each of the zones.
Outside the interaction region ($\alpha=I,III$), 
\begin{eqnarray}
\left(
\begin{array}{c}
g_\alpha(x)\\
e_\alpha(x)\\
\dot{g}_\alpha(x)\\
\dot{e}_\alpha(x)
\end{array}\right)=M_0(x)
\left(
\begin{array}{c}
a_\alpha\\
b_\alpha\\
c_\alpha\\
d_\alpha
\end{array}\right),
\end{eqnarray}
and inside the field ($\alpha=II$)
\begin{eqnarray}
\left(
\begin{array}{c}
g_\alpha(x)\\
e_\alpha(x)\\
\dot{g}_\alpha(x)\\
\dot{e}_\alpha(x)
\end{array}\right)=M_b(x,\phi_1)
\left(
\begin{array}{c}
a_\alpha\\
b_\alpha\\
c_\alpha\\
d_\alpha
\end{array}\right), 
\end{eqnarray}
where the dot represents derivative with respect to $x$. 
The $M$ matrices are explicitly given by
\begin{eqnarray}
M_0(x) &=&  \left(
\begin{array}{cccc}
e^{ikx} & e^{-ikx} & 0 & 0 \\
0 & 0 & e^{iqx} & e^{-iqx} \\
ik e^{ikx} & -ik e^{-ikx} & 0 & 0 \\
0 &  0 & iq e^{iqx} & -iq e^{-iqx}
\end{array} \right) ,
\nonumber
\end{eqnarray}
\begin{widetext}
\begin{eqnarray}
M_b(x,\phi) &=& \left(
\begin{array}{cccc}
e^{ik_+ x} & e^{-ik_+ x} & e^{ik_- x} & e^{-ik_- x} \\
\frac{2\lambda_+}{\Omega} e^{-i\phi} e^{ik_+ x} & \frac{2\lambda_+}{\Omega} e^{-i\phi} e^{-ik_+ x} & \frac{2\lambda_-}{\Omega} e^{-i\phi} e^{ik_- x} & \frac{2\lambda_-}{\Omega} e^{-i\phi} e^{-ik_- x} \\
ik_+ e^{ik_+ x} & -ik_+ e^{-ik_+ x} & ik_- e^{ik_- x} & -ik_- e^{-ik_- x} \\
ik_+ \frac{2\lambda_+}{\Omega} e^{-i\phi} e^{ik_+ x} &  -ik_+ \frac{2\lambda_+}{\Omega} e^{-i\phi} e^{-ik_+ x} & ik_- \frac{2\lambda_-}{\Omega} e^{-i\phi} e^{ik_- x} & -ik_- \frac{2\lambda_-}{\Omega} e^{-i\phi} e^{-ik_- x} 
\end{array}\right).
\end{eqnarray}
\end{widetext}
With this notation, the matching conditions at $x=x_1$ and $x=x_2$ can be written as
\begin{eqnarray}
M_0(x_1)\boldsymbol{v}_{_I}=M_b(x_1,\phi_1)\boldsymbol{v}_{_{II}},\\
M_b(x_2,\phi_1)\boldsymbol{v}_{_{II}}=M_0(x_2)\boldsymbol{v}_{_{III}}.
\end{eqnarray}
Eliminating $\boldsymbol{v}_{_{II}}$ from the system above, we end up with a transfer matrix $\textsf{T}(x_1,x_2,\phi_1)$ which connects
the amplitudes of both sides, 
\begin{equation}
\boldsymbol{v}_{_I}=\textsf{T}(x_1,x_2,\phi_1)\boldsymbol{v}_{_{III}},
\end{equation}
defined by
\begin{equation}
\textsf{T}(x_1,x_2,\phi)=M_0(x_1)^{-1}M_b(x_1,\phi)M_b(x_2,\phi)^{-1}M_0(x_2).
\label{tram}
\end{equation}
%
\subsection{Phase dependence}
The explicit dependence of the (one laser) transfer matrix on the laser phase $\phi$ 
(we drop the laser index $n$) is as follows 
\begin{eqnarray}
\textsf{T}(x_1,x_2,\phi)=\left(\begin{array}{cccc}
\tilde{\textsf{T}}_{11}&\tilde{\textsf{T}}_{12}&e^{i\phi}
\tilde{\textsf{T}}_{13}&e^{i\phi}\tilde{\sa{T}}_{14}\\
\tilde{\sa{T}}_{21}&\tilde{\sa{T}}_{22}&e^{i\phi}\tilde{\sa{T}}_{23}&e^{i\phi}
\tilde{\sa{T}}_{24}\\
e^{-i\phi}\tilde{\sa{T}}_{31}&e^{-i\phi}\tilde{\sa{T}}_{32}&\tilde{\sa{T}}_{33}&
\tilde{\sa{T}}_{34}\\
e^{-i\phi}\tilde{\sa{T}}_{41}&e^{-i\phi}\tilde{\sa{T}}_{42}&
\tilde{\sa{T}}_{43}&\tilde{\sa{T}}_{44}
                      \end{array}\right),
\nonumber\end{eqnarray}
where the tildes represent the phase-free form
of the amplitudes, i. e., $\tilde T_{ij}=T_{ij}(\phi=0)$.
\subsection{Multiple laser fields}
Clearly we may repeat step by step the operations above for the second and
third laser. The results are formally the same, except for the 
substitution of the 
matching points and the laser phase.  
We may then write
\begin{eqnarray}
\boldsymbol{v}_{_{I}}=\textsf{T}(x_1,x_2,\phi_1)\boldsymbol{v}_{_{III}},\\
\boldsymbol{v}_{_{III}}=\textsf{T}(x_3,x_4,\phi_2)\boldsymbol{v}_{_V},\\
\boldsymbol{v}_{_V}=\textsf{T}(x_5,x_6,\phi_3)\boldsymbol{v}_{_{VII}}, 
\end{eqnarray}
and relate the waves on the extremes by
\begin{equation}
\boldsymbol{v}_{_I}=\textsf{T}(x_1,x_2,\phi_1)\textsf{T}(x_3,x_4,\phi_2)\textsf{T}(x_5,x_6,\phi_3)\boldsymbol{v}_{_{VII}}.
\end{equation}
%
\section{Recurrence Relations}
\label{recurrence}
%
%
Consider, for the $n$th laser located between $x_i$
and $x_f$, the following ``elementary'' scattering boundary conditions
corresponding to incidence of a wave in one channel from
left or right:   
\begin{itemize} 
\item Left incoming, ground state:
\begin{eqnarray}
\left(\begin{array}{c}
1\\
(r_n)_{gg}^l\\
0\\
(r_n)_{ge}^l
      \end{array}\right)=\textsf{T}^{(n)}(x_i,x_f)\left(\begin{array}{c}
(t_{n})_{gg}^l\\
0\\
(t_{n})_{ge}^l\\
0
      \end{array}\right).
\end{eqnarray}
\item Left incoming, excited state:
\begin{eqnarray}
\left(\begin{array}{c}
0\\
(r_n)_{eg}^l\\
1\\
(r_n)_{ee}^l
      \end{array}\right)=\textsf{T}^{(n)}(x_i,x_f)\left(\begin{array}{c}
(t_{n})_{eg}^l\\
0\\
(t_{n})_{ee}^l\\
0
      \end{array}\right).
\end{eqnarray}
\item Right incoming, ground state:
\begin{eqnarray}
\left(\begin{array}{c}
0\\
(t_{n})_{gg}^r\\
0\\
(t_{n})_{ge}^r
      \end{array}\right)=
\textsf{T}^{(n)}(x_i,x_f)\left(\begin{array}{c}
(r_n)_{gg}^r\\
1\\
(r_n)_{ge}^r\\
0
      \end{array}\right).
\end{eqnarray}
\item Right incoming, excited state:
\begin{eqnarray}
\left(\begin{array}{c}
0\\
(t_{n})_{eg}^r\\
0\\
(t_{n})_{ee}^r
      \end{array}\right)=
\textsf{T}^{(n)}(x_i,x_f)\left(\begin{array}{c}
(r_n)_{eg}^r\\
0\\
(r_n)_{ee}^r\\
1
      \end{array}\right).
\end{eqnarray}
\end{itemize}
Thus, for each laser we have a system of $16$ equations which can be 
solved to 
%
%
%
%
%
%
%
give the 
transfer matrix $\T^{(n)}$ elements as a function of the single field scattering amplitudes
$(r_\alpha)_{ij}^{r,l}$, or the other way around, the single field scattering amplitudes in terms of the 
transfer matrix elements. Combined with Eq. (\ref{tram}), 
this provides explicit, exact  expressions 
for the scattering amplitudes.  
%
\subsection{$\textsf{T}^{(n)}_{ij}$ as a function of $(r_n)_{i,j}^{l,r}$ and $(t_n)_{i,j}^{l,r}$}
We have dropped the $n$ index of the laser for simplicity.
\begin{eqnarray}
\sa{T}_{11}&=&{t_{ee}^l}/{f}\nonumber\\
\sa{T}_{12}&=&({r_{ge}^r t_{eg}^l-r_{gg}^r t_{ee}^l})/{f}\nonumber\\
\sa{T}_{13}&=&{-t_{eg}^l}/{f}\nonumber\\
\sa{T}_{14}&=&({r_{ee}^r t_{eg}^l-r_{eg}^r t_{ee}^l})/{f}\nonumber\\
\sa{T}_{21}&=&({r_{gg}^l t_{ee}^l - r_{eg}^l t_{ge}^l})/{f}\nonumber\\
\sa{T}_{22}&=&t_{gg}^r
-\frac{r_{gg}^l r_{gg}^r t_{ee}^l - r_{gg}^l r_{ge}^r t_{eg}^l - r_{eg}^l r_{gg}^r t_{ge}^l+
r_{eg}^l r_{ge}^r t_{gg}^l}{f}\nonumber\\
\sa{T}_{23}&=&({r_{eg}^l t_{gg}^l-r_{gg}^l t_{eg}^l})/{f}\nonumber\\
\sa{T}_{24}&=&t_{eg}^r
-\frac{r_{gg}^l r_{eg}^r t_{ee}^l - r_{gg}^l r_{ee}^r t_{eg}^l - r_{eg}^l r_{eg}^r t_{ge}^l + r_{eg}^l r_{ee}^r t_{gg}^l}{f}\nonumber\\
\sa{T}_{31}&=&{-t_{ge}^l}/{f}\nonumber\\
\sa{T}_{32}&=&({r_{gg}^r t_{ge}^l - r_{ge}^r t_{gg}^l})/{f}\nonumber\\
\sa{T}_{33}&=&{t_{gg}^l}/{f}\nonumber\\
\sa{T}_{34}&=&({r_{eg}^r t_{ge}^l - r_{ee}^r t_{gg}^l})/{f}\nonumber\\
\sa{T}_{41}&=&({r_{ge}^l t_{ee}^l - r_{ee}^l t_{ge}^l})/{f}\nonumber\\
\sa{T}_{42}&=&t_{ge}^r
-\frac{r_{ge}^l r_{gg}^r t_{ee}^l - r_{ge}^l r_{ge}^r t_{eg}^l - r_{ee}^l r_{gg}^r t_{ge}^l +
r_{ee}^l r_{ge}^r t_{gg}^l}{f}\nonumber\\
\sa{T}_{43}&=&({r_{ee}^l t_{gg}^l-r_{ge}^l t_{eg}^l})/{f}\nonumber\\
\sa{T}_{44}&=&t_{ee}^r
-\frac{r_{ge}^l r_{eg}^r t_{ee}^l - r_{ge}^l r_{ee}^r t_{eg}^l - r_{ee}^l r_{eg}^r t_{ge}^l +r_{ee}^l r_{ee}^r t_{gg}^l}{f}\nonumber
\end{eqnarray}
with the common denominator $f$ defined by
\begin{equation}
f=t_{ee}^l t_{gg}^l-t_{eg}^l t_{ge}^l.
\end{equation}
\subsection{$(r_n)_{i,j}^{l,r}$ and $(t_n)_{i,j}^{l,r}$ as a function of $\sa{T}^{(n)}_{ij}$}
We have dropped the $n$ index of the laser for simplicity.
\begin{eqnarray}
r^l_{gg} &=& -({-\sa{T}_{23} \sa{T}_{31} + \sa{T}_{21} \sa{T}_{33}})/{F}\nonumber\\ 
r^l_{ge} &=& ({-\sa{T}_{33} \sa{T}_{41} + \sa{T}_{31} \sa{T}_{43}})/{F}\nonumber\\ 
r^l_{eg} &=& -({-\sa{T}_{13} \sa{T}_{21} + \sa{T}_{11} \sa{T}_{23}})/{F}\nonumber\\ 
r^l_{ee} &=& -({-\sa{T}_{13} \sa{T}_{41} + \sa{T}_{11} \sa{T}_{43}})/{F}\nonumber\\ 
r^r_{gg} &=& ({-\sa{T}_{13} \sa{T}_{32} + \sa{T}_{12} \sa{T}_{33}})/{F}\nonumber\\ 
r^r_{ge} &=& -({\sa{T}_{12} \sa{T}_{31} - \sa{T}_{11} \sa{T}_{32}})/{F}\nonumber\\ 
r^r_{eg} &=& ({\sa{T}_{14} \sa{T}_{33} - \sa{T}_{13} \sa{T}_{34}})/{F}\nonumber\\ 
r^r_{ee} &=& -({\sa{T}_{14} \sa{T}_{31} - \sa{T}_{11} \sa{T}_{34}})/{F}\nonumber\\ 
t^l_{gg} &=& -{\sa{T}_{33}}/{F}\nonumber\\ 
t^l_{ge} &=& {\sa{T}_{31}}/{F}\nonumber\\ 
t^l_{eg} &=& {\sa{T}_{13}}/{F}\nonumber\\ 
t^l_{ee} &=& -{\sa{T}_{11}}/{F}\label{tr1}\\
\end{eqnarray}
\begin{widetext}
\begin{eqnarray}
t^r_{gg} &=& -\frac{-\T_{13} \T_{22} \T_{31} + \T_{12} \T_{23} \T_{31} +   \T_{13} \T_{21} \T_{32} - \T_{11} \T_{23} \T_{32} - \T_{12} \T_{21} \T_{33} + \T_{11} \T_{22} \T_{33}}{
  F}\nonumber\\ 
t^r_{ge} &=& -\frac{ \T_{13} \T_{32} \T_{41} - \T_{12} \T_{33} \T_{41} - \T_{13} \T_{31} \T_{42} + \T_{11} \T_{33} \T_{42} + 
   \T_{12} \T_{31} \T_{43} - \T_{11} \T_{32} \T_{43}}{F}\nonumber\\ 
t^r_{eg} &=& -\frac{ \T_{14} \T_{23} \T_{31} - \T_{13} \T_{24} \T_{31} - \T_{14} \T_{21} \T_{33} + \T_{11} \T_{24} \T_{33} + 
   \T_{13} \T_{21} \T_{34} - \T_{11} \T_{23} \T_{34}}{F}\nonumber\\ 
t^r_{ee} &=& -\frac{-\T_{14} \T_{33} \T_{41} + \T_{13} \T_{34} \T_{41} + 
   \T_{14} \T_{31} \T_{43} - \T_{11} \T_{34} \T_{43} - \T_{13} \T_{31} \T_{44} + \T_{11} \T_{33} \T_{44}}{
  F}
\label{tr2}  
\end{eqnarray}
\end{widetext}
where the common denominator $F$ is given by
\begin{equation}
F=\T_{13}\T_{31}-\T_{11}\T_{33}
\end{equation}
%
%
%
\section{Explicit expressions of the quantum scattering amplitudes}
\label{explicit_ampl_appendix}
We give here explicit expressions for the amplitudes in Eq. (\ref{tge_26paths}). 
%
\subsection{$4$ direct paths}
\label{4_paths_app}
There are four possible paths leading to an excited atom with no reflection. 
These are the corresponding amplitudes,
\begin{eqnarray}
A_1^q&=&(t_1)_{ge}^l (t_2)_{ee}^l (t_3)_{ee}^l,\\
A_2^q&=&(t_1)_{ge}^l (t_2)_{eg}^l (t_3)_{ge}^l,\\
A_3^q&=&(t_1)_{gg}^l (t_2)_{ge}^l (t_3)_{ee}^l,\\
A_4^q&=&(t_1)_{gg}^l (t_2)_{gg}^l (t_3)_{ge}^l.
\end{eqnarray}
%
\subsection{$22$ paths with two reflections}
\label{22_paths_app}
The quantum amplitudes $B_i^q$ 
for all the $22$ possible paths leading to a transmitted
excited atom including two reflections, provided that the perfect $\pi$-pulse at the
second laser flips the atomic state are explicitly given by 
\begin{eqnarray}
B_1^q&=&(t_1)_{gg}^l  (r_2)_{gg}^l  (r_1)_{gg}^r (t_2)_{ge}^l (t_3)_{ee}^l,\nonumber\\ 
B_2^q&=&(t_1)_{gg}^l  (r_2)_{gg}^l  (r_1)_{ge}^r  (t_2)_{eg}^l (t_3)_{ge}^l,\nonumber\\ 
B_3^q&=&(t_1)_{gg}^l  (t_2)_{ge}^l  (r_3)_{eg}^l  (r_2)_{gg}^r  (t_3)_{ge}^l,\nonumber\\ 
B_{4}^q&=&(t_1)_{ge}^l  (t_2)_{eg}^l   (r_3)_{gg}^l  (r_2)_{gg}^r   (t_3)_{ge}^l,\nonumber\\ 
B_5^q&=& (t_1)_{gg}^l  (r_2)_{ge}^l  (r_1)_{eg}^r  (t_2)_{ge}^l (t_3)_{ee}^l,\nonumber\\ 
B_6^q&=&(t_1)_{gg}^l  (r_2)_{ge}^l  (r_1)_{ee}^r  (t_2)_{eg}^l (t_3)_{ge}^l,\nonumber\\ 
B_7^q&=&(t_1)_{gg}^l  (t_2)_{ge}^l  (r_3)_{eg}^l  (r_2)_{ge}^r   (t_3)_{ee}^l,\nonumber\\ 
B_{8}^q&=&(t_1)_{ge}^l  (t_2)_{eg}^l  (r_3)_{gg}^l  (r_2)_{ge}^r (t_3)_{ee}^l,\nonumber\\ 
B_{9}^q&=&(t_1)_{ge}^l  (r_2)_{eg}^l  (r_1)_{gg}^r  (t_2)_{ge}^l (t_3)_{ee}^l,\nonumber\\ 
B_{10}^q&=&(t_1)_{ge}^l  (r_2)_{eg}^l  (r_1)_{ge}^r   (t_2)_{eg}^l (t_3)_{ge}^l,\nonumber\\ 
B_{11}^q&=&(t_1)_{gg}^l  (t_2)_{ge}^l  (r_3)_{ee}^l  (r_2)_{ee}^r   (t_3)_{ee}^l,\nonumber\\ 
B_{12}^q&=&(t_1)_{gg}^l  (t_2)_{ge}^l  (r_3)_{ee}^l  (t_2)_{eg}^r  (r_1)_{gg}^r  (t_2)_{ge}^l (t_3)_{ee}^l,\nonumber\\ 
B_{13}^q&=&(t_1)_{gg}^l  (t_2)_{ge}^l  (r_3)_{eg}^l  (t_2)_{ge}^r  (r_1)_{eg}^r  (t_2)_{ge}^l (t_3)_{ee}^l,\nonumber\\ 
B_{14}^q&=&(t_1)_{gg}^l  (t_2)_{ge}^l  (r_3)_{ee}^l  (t_2)_{eg}^r  (r_1)_{ge}^r (t_2)_{eg}^l  (t_3)_{ge}^l,\nonumber\\ 
B_{15}^q&=&(t_1)_{gg}^l  (t_2)_{ge}^l  (r_3)_{eg}^l  (t_2)_{ge}^r  (r_1)_{ee}^r (t_2)_{eg}^l  (t_3)_{ge}^l,\nonumber\\ 
B_{16}^q&=&(t_1)_{ge}^l (t_2)_{eg}^l (r_3)_{ge}^l  (r_2)_{ee}^r (t_3)_{ee}^l,\nonumber\\ 
B_{17}^q&=&(t_1)_{ge}^l  (r_2)_{ee}^l  (r_1)_{eg}^r   (t_2)_{ge}^l (t_3)_{ee}^l,\nonumber\\ 
B_{18}^q&=&(t_1)_{ge}^l  (t_2)_{eg}^l  (r_3)_{ge}^l  (t_2)_{eg}^r  (r_1)_{gg}^r  (t_2)_{ge}^l  (t_3)_{ee}^l,\nonumber\\ 
B_{19}^q&=&(t_1)_{ge}^l  (t_2)_{eg}^l  (r_3)_{gg}^l  (t_2)_{ge}^r   (r_1)_{eg}^r  (t_2)_{ge}^l  (t_3)_{ee}^l,\nonumber\\ 
B_{20}^q&=&(t_1)_{ge}^l  (r_2)_{ee}^l   (r_1)_{ee}^r (t_2)_{eg}^l (t_3)_{ge}^l,\nonumber\\ 
B_{21}^q&=&(t_1)_{ge}^l  (t_2)_{eg}^l  (r_3)_{ge}^l  (t_2)_{eg}^r  (r_1)_{ge}^r  (t_2)_{eg}^l  (t_3)_{ge}^l,\nonumber\\ 
B_{22}^q&=&(t_1)_{ge}^l  (t_2)_{eg}^l  (r_3)_{gg}^l  (t_2)_{ge}^r  (r_1)_{ee}^r  (t_2)_{eg}^l   (t_3)_{ge}^l.
\end{eqnarray}
%
\subsection{Phase dependence of the scattering amplitudes}
\label{phase_dep_app}
The dependence of each of the path amplitudes on the laser phases 
are easily  obtained from the  $2$-channel recurrence relations and the transfer matrix formalism. If the phase-free amplitudes (for all $\phi_n=0$)
are denoted by tildes, we have 
\begin{eqnarray}
A_1^q&=&e^{-i\phi_1}\tilde A_1^q,\nonumber\\
A_2^q&=&e^{-i(\phi_1-\phi_2+\phi_3)}\tilde A_2^q,\nonumber\\
A_3^q&=&e^{-i\phi_2}\tilde A_3^q,\nonumber\\
A_4^q&=&e^{-i\phi_3}\tilde A_4^q,\nonumber\\
B_1^q&=&e^{-i\phi_2}\tilde B_1^q,\nonumber\\
B_2^q&=&e^{-i(\phi_1-\phi_2+\phi_3)}\tilde B_2^q,\nonumber\\
B_3^q&=&e^{-i\phi_2}\tilde B_3^q,\nonumber\\
B_4^q&=&e^{-i(\phi_1-\phi_2+\phi_3)}\tilde B_4^q.
\end{eqnarray}


\end{document}